\documentclass[pra,twocolumn,showpacs,preprintnumbers,amsmath,amssymb]{revtex4}
\usepackage{graphicx}
\usepackage{dcolumn}
\usepackage{bm}
\usepackage{latexsym}
\usepackage{amsfonts}
\usepackage{xcolor,MnSymbol}

\begin{document}
\title{The demixing of the two-component fermions in optical lattices under a spin-dependent external potential}
\author{A-Hai Chen}
\author{Gao Xianlong}
\email{gaoxl@zjnu.edu.cn}
\affiliation{Department of Physics, Zhejiang Normal University, Jinhua 321004, China}

\date{\today}

\begin{abstract}
The demixing of two-component fermions in optical lattices under a spin-dependent external potential is investigated using the density-matrix renormalization group method. The influence of on-site interactions (u) and the ratio of the spin-dependent external potentials ($\gamma$) on the demixing is discussed. The $\gamma-u$ state diagram is numerically mapped out, from which we distinguish the phase-mixed region from the phase-separated one. A minimum point appears in the phase boundary. On the right part of the point an insulating core is formed in the spin-up densities during the demixing, while on the left one only metallic phase exists.
\end{abstract}
\pacs{03.75.Ss,67.85.Lm,71.10.Pm}
\maketitle

\section{Introduction}
\label{intro}
Cold-atom experiments provide a clean and versatile platform for simulating interacting quantum many-body systems and investigating new quantum phases~\cite{optical,advinphyreivew2007}. In a system of cold atomic mixtures~\cite{Das}, spin-dependent external potentials can be introduced by the different masses or the hyperfine states of the constituent~\cite{spindependent_exp}.
Spatial phase separation (i.e., demixing) of the components happens at repulsive interactions, which on the one hand will diminish the cooling efficiency in a sympathetic cooling procedure~\cite{FBM_demix1}. On the other hand, at stronger repulsive interactions and/or tighter confinement, this procedure provides a way to demix the different atomic species~\cite{trap-imbalance2,trap-imbalance1,pra-spin}.

Different types of mixtures in optical lattices~\cite{optical-hubbard1} have been studied, including boson mixtures~\cite{BM_demix2}, fermion mixtures of same or different species~\cite{FM2,FM_demix1,attractivemix}, which can be well modelled by a lattice model interacting through a Hubbard-type term with confining potentials. By detuning asymmetrically the laser frequencies with respect to the two hyperfine states~\cite{trap-imbalance1}, or by trapping the two different atoms of unequal masses, one realizes the different types of spin-dependent external potentials~\cite{spind-hubbard2,spind-hubbard3,spind-hubbard4}.

For a better understanding of the interplay of the spin-dependent external potentials and the repulsive interaction on the process of demixing, it is essential to have a complete state diagram, from which one can easily find the optimal parameters to realize the demixing or to control the cooling efficiency. In this paper, we study the demixing of two fermion species (taking as pseudospins) of same masses in optical lattices with spin-dependent external potentials.

This brief report is organized as follows. In Sec. II the modified Hubbard model under a spin-dependent external potential is introduced. Section III contains the main numerical results. Finally, we summarize the paper.

\section{The model}
\label{model}

A two-component Fermi gas in optical lattices with a spin-dependent external potential can be described by a modified one-dimension (1D) Fermi-Hubbard model as
\begin{eqnarray}\label{eq:hubbard}
\hat {H}&=&-t\sum_{i,\sigma}^{}({\hat
c}^{\dagger}_{i\sigma}{\hat c}_{i+1\sigma}+{H}.{c}.)+
U\sum_{i}{\hat n}_{i\uparrow}{\hat n}_{i\downarrow}\nonumber\\
&&+\sum_{i,\sigma}^{}V_\sigma\left[i-(L-1)/2\right]^2{\hat n}_{i\sigma}\,,
\end{eqnarray}
where $t$ is the hopping parameter between nearest neighbors and $U$ the on-site repulsive interaction. The strength of the spin-dependent external potential is $V_{\sigma}$. The two components fermions are represented by pseudospin $\sigma=\uparrow, \downarrow$. $L$ is the length of the lattice, which is chosen to be long enough to keep the atomic densities at edges to be zero. $\hat{c}^{\dagger}_{i\sigma}$ (${\hat c}_{i\sigma}$) is the creation (annihilation) operator on site $i$ with spin $\sigma$. The density operator for spin $\sigma$ is ${\hat n_{i\sigma}}$.

By taking $t$ as the unit of energy, we rewrite the Hamiltonian as
\begin{eqnarray}\label{eq:hubbard}
\tilde{\hat {H}}&=&- \sum_{i,\sigma}^{}({\hat
c}^{\dagger}_{i\sigma}{\hat c}_{i+1\sigma}+{H}.{c}.)+
u\sum_{i=1}^{L}\,{\hat n}_{i\uparrow}{\hat n}_{i\downarrow}\nonumber\\
&&+\sum_{i}^{}v_{\downarrow}\left[i-(L-1)/2\right]^2(\gamma{\hat n}_{i\uparrow}+{\hat n}_{i\downarrow})\,,
\end{eqnarray}
with rescaled $u=U/t$ and $v_{\sigma}=V_{\sigma}/t$. $\gamma=v_\uparrow/v_\downarrow$ is the ratio between the spin-up and spin-down external potentials. For convenience, we assume that $v_\uparrow \ge v_\downarrow$. Thus, $\gamma \ge 1 $. The number of fermions of each species is $N_\sigma=\sum_i {n}_{i\sigma}=\sum_i\langle \hat{n}_{i\sigma}\rangle$ with ${n}_{i\sigma}$ the spin-resolved density and the total number of fermions $N_f=N_\uparrow+N_\downarrow$. The polarization $P=(N_\uparrow-N_\downarrow)/N_f$. The local magnetization is defined as $m_i=({n}_{i\uparrow}-{n}_{i\downarrow})/2$. When $\gamma=1$, the model recovers the mostly studied Hubbard model in the presence of a harmonic trap~\cite{Rigol,Rigol2004,Liu,Heiselberg,Campo}. We are interested in the effects induced by the spin-dependent external potentials. For atomic mixture of different species in optical lattices formed by lasers of wavelength $1024 nm$ the ratio of the trapping potential $\gamma$ can reach $2.3$ for fermionic mixtures of $^{6}{\rm Li}$ and $^{40}{\rm K}$, and $2.5$ for $^6{\rm Li}$ and $^{87}{\rm Sr}$. Larger $\gamma$ is hard to achieve at the present experiment but conceivable.

\section{NUMERICAL RESULTS}
\label{result}
We study in the following the effects of $u$ and $\gamma$ on the phase separation.
Our simulation is done by using the density-matrix renormalization group (DMRG)~\cite{DMRG,alps}. In principle the number of states kept are 300 per block and 30 sweeps are used (when checking the numerical accuracy, up to 1000 states per block and 50 sweeps are used), with a truncation error in the final DMRG step usually smaller than $10^{-7}$.
Without mentioning in the text, we will choose the spin-unpolarized system with fixed particle number $N_f=40$ and $v_\downarrow=10^{-3}$.

\begin{figure}
\begin{center}
\resizebox{1.0\columnwidth}{!}{%
  \includegraphics{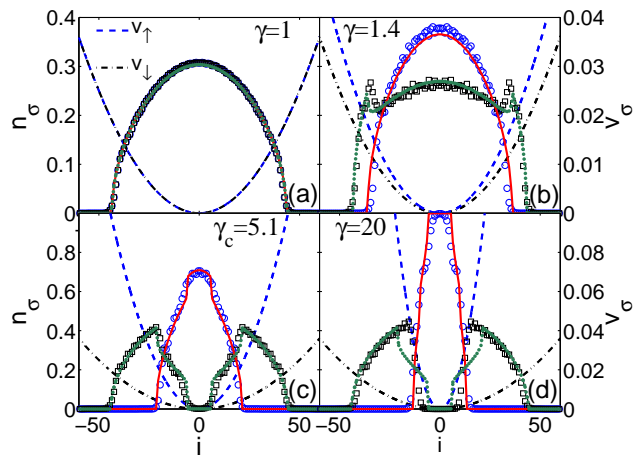}}
\caption{(Color online)  The density distribution of the two fermionic components under a spin-dependent harmonic potential $v_{\sigma}$ (dashed line for $v_{\uparrow}$ and dot-dashed for $v_{\downarrow}$). The parameters are $u=5$, $N_f =40$, and $v_\downarrow=0.001$. The DMRG data (open circle for $n_\uparrow$ and open square for $n_\downarrow$) are compared against the LSDA ones (solid line for $n_\uparrow$ and dotted line for $n_\downarrow$). Panel (a): The two spin components have same distribution due to the spin-independent external potential at $\gamma=1$. Panel (b): A part of spin-down atoms is forced out of system center with larger $\gamma=1.4$. Panels (a) and (b) belong to a PM region. In panel (c), at the critical point $\gamma_c=5.1$ when there is no spin-down atoms left at the center of the system, a PS region starts to form. Panel (d): A real PS region is formed where the two components are demixed at $\gamma=20$.}
\label{fig:one}
\end{center}
\end{figure}

We illustrate the physics induced by $\gamma$ in Fig.~\ref{fig:one}. In panel (a), due to the same external potential, the spin-up and spin-down atoms have the same densities and accumulate at the potential minimum. With the increase of $\gamma$, the spin-up atoms suffer from stronger trapping potential, and part of the spin-down atoms are forced out of the system center, as the result of the competition between the potential energy and the interaction energy, which is shown in panel (b). Panel (b) including (a) is the case where the spin-up and spin-down atom mixture coexists in the center of the trap and forms a phase-mixed (PM) region. In panel (c), spin-up and spin-down atoms are becoming phase-separated at a critical $\gamma_c=5.1$, decided when the local occupation for spin-down atoms disappears at the center (Numerically, we get the critical value of $\gamma_c$ when $n_{i\downarrow} \approx 10^{-3}$ with $i$ ranging over the two central points~\cite{footnote}.).
Beyond the critical $\gamma_c$, the spin-up and spin-down atomic clouds have diminished spatial overlap and are phase-separated~\cite{trap-imbalance1,pra-spin}.
A completely phase-separated case is shown at $\gamma\gg\gamma_c$ in panel (d). Panel (c) including (d) defines a phase-separated (PS) region. For comparison, in Fig.~\ref{fig:one}, we present also the results based on the local spin-density approximation (LSDA) by using a recent parametrized energy functional for the 1D Hubbard model~\cite{LSDA}. In the LSDA, the energy density of the inhomogeneous system is approximated at each spatial position by the one corresponding to a uniform gas at the local value of the density. We find that the LSDA gives qualitatively the same results as those of DMRG. However, the performance of the LSDA scheme at weaker interactions deteriorates with decreasing particle number where the hopping kinetic energy processes. As a result, the regions close to the edges of the trap and the phase-separation areas are also those where the LSDA is less accurate.

\begin{figure}
\begin{center}
\resizebox{0.75\columnwidth}{!}{%
  \includegraphics{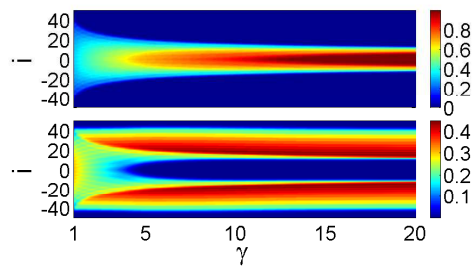}}
\caption{(Color online) A contour plot for the influence of $\gamma$ on the demixing of the two components. The parameters are the same as in Fig.~\ref{fig:one} but with continuously varying $\gamma$. Top panel: spin-up density ($n_\uparrow$) and down panel: spin-down density ($n_\uparrow$) are shown, respectively. With the increase of the $\gamma$, spin-up atoms become more accumulated and the spin-down atoms are forced out of the center of the system gradually due to the stronger repulsive interactions between them.}
\label{fig:gamma}
\end{center}
\end{figure}
\begin{figure}
\begin{center}
\resizebox{0.75\columnwidth}{!}{%
  \includegraphics{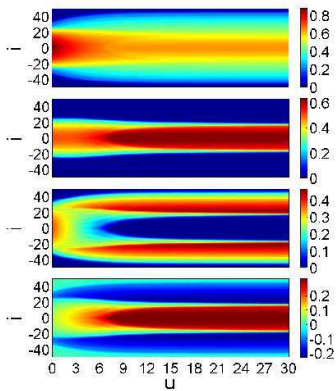}}
\caption{(Color online) The influence of $u$ on the demixing of the two components. The system consists of $N_\uparrow=20$ and $N_\downarrow=20$ fermions. The spin-down trap strength is $10^{-3}$ and $\gamma=3$. From top to bottom, the total density distribution ($n$), spin-up density ($n_\uparrow$), spin-down density ($n_\uparrow$), and the magnetization (m) are shown, respectively.}
\label{fig:u}
\end{center}
\end{figure}

To have a complete understanding on the effect of $\gamma$ on the phase separation, in Fig. \ref{fig:gamma}, the contour plot is shown for the spin-up density $n_\uparrow$ (up panel) and spin-down density $n_\downarrow$ (down panel) with $1\le \gamma\le 20$.
When $\gamma>12$, an insulating core of fully polarized (spin-up) atoms ($n_{i\uparrow}=1$) is formed due to the strong confinement, which is incompressible and originates from the Pauli exclusion principle. As a result, the magnetization tends to 0.5. Around $\gamma\sim 20$, stable density distributions are formed and no more changes can be observed by further increasing $\gamma$.

The effects of the repulsive interaction $u$ (with fixed $\gamma=3$) on the phase separation are given in Fig. \ref{fig:u}. For a non-interacting system ($u=0$), the two components keep mixed. The spin-down atoms are forced out of the center due to the strong repulsive interaction between the components when increasing $u$. The system reaches the critical PS point at $u_c=8$. The two fermion components are fully separated around $u\sim 15$ at which the density distributions become stable.

\begin{figure}
\begin{center}
\resizebox{1.0\columnwidth}{!}{%
  \includegraphics{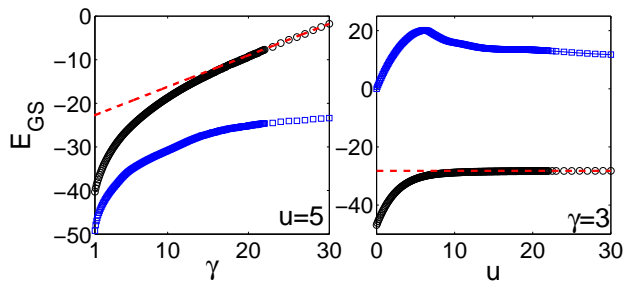}}
\caption{(Color online) (Left panel) Ground-state energy $E_{GS}$ (open circle) as a function of $\gamma$. The parameters are the same as Fig.~\ref{fig:gamma}. The energy without the contribution from the spin-down external potential $E_{A}$ (open square) is also plotted. The dash line is the linear fit for $E_{GS}$ with $E_{GS}=-24+0.73\gamma$.
(Right panel) Ground-state energy $E_{GS}$ (open circle) as a function of $u$. The parameters are the same as Fig.~\ref{fig:u}. The interaction energy $E_u$ is also shown (open square), which starts from zero when there is no interaction, reaches a maximum at $U=5$, and goes to a constant when the two components segregate almost completely. The dash line is a constant at $E=-28.22$.}
\label{fig:energy}
\end{center}
\end{figure}

The interplay of $\gamma$ and $u$ segregates the two components. The spin-up atomic cloud that suffers from a stronger external potential tends to accumulate at the center. As a result, the spin-down atoms are repelled from the center due to a larger repulsion, which decreases the interaction energy but increases the external potential energy. The equilibrium density distribution is the result of the balance between these two opposite effects. The shape of the density profiles becomes stable at further increasing $\gamma$ and $u$. To understand this point, we plot in Fig. \ref{fig:energy} the energies as functions of $\gamma$ and $u$, respectively. In the left panel, we study the ground-state energy
\begin{eqnarray}\label{hubbardr}
E_{GS}=T+uD+E^{\downarrow}_{ext}+\gamma E^{\uparrow}_{ext}\,,
\end{eqnarray}
and the energy except the last term $E_A=T+uD+E^{\downarrow}_{ext}$ as a function of the ratio $\gamma$.
Here $T$ is the kinetic energy, $D$ the double occupancy $D=\sum_{i=1}^{L}\langle{\hat n}_{i\uparrow}{\hat n}_{i\downarrow}\rangle$, and the respective
external potential energy is $E^{\sigma}_{ext}=\sum_{i,\sigma}^{}v_\downarrow\left[i-(L-1)/2\right]^2\langle {\hat n}_{i\sigma} \rangle$.
We notice that at $\gamma \gtrsim 20$, $E_{GS}$ becomes a linear function of $\gamma$: $E_{GS}=-24+0.73\gamma$ and $E_A$ goes to a constant $-24$, which indicates that the densities are becoming stable and the only contribution to the energy comes from the last term linear to $\gamma$. In the right panel, we study the ground-state energy $E_{GS}$ and the interaction energy $E_u=uD$ as a function of the interaction strength $u$.
We can see that at $u\gtrsim 22$, $E_{GS}$ and $E_u$ become constant, which indicates that the two components are demixed almost completely except the tails of the clouds. Under the stable density distributions, the on-site interaction contributes a constant energy.
\begin{figure}
\begin{center}
\resizebox{1.0\columnwidth}{!}{%
  \includegraphics{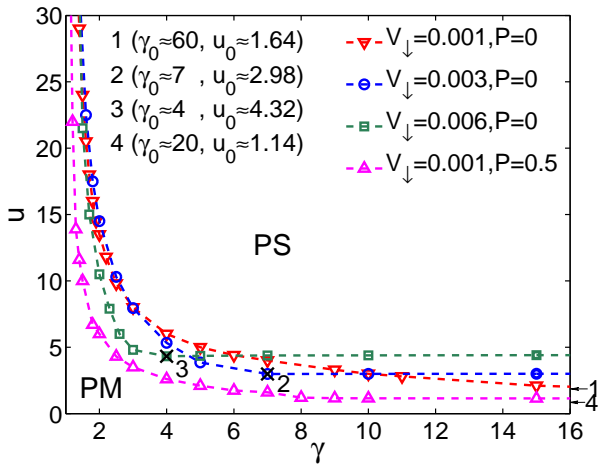}}
\caption{(Color online) The $\gamma-u$ state diagram for different systems. For the unpolarized systems, the phase boundaries for three different $v_\downarrow=0.001, 0.003$, and $0.006$ are shown by down-triangle, circle, and square, respectively. For the polarized system for $v_\downarrow=0.001$ ($P=0.5$), it is shown by the up-triangle. The number 2 and 3 are the corresponding minimum position in the phase boundary, while 1 and 4 mark only the corresponding minimum positions in the $u$-axis.
The dashes lines are just a guide for eyes.}
\label{fig:phase}       
\end{center}
\end{figure}
\begin{figure}
\begin{center}
\resizebox{1.0\columnwidth}{!}{%
  \includegraphics{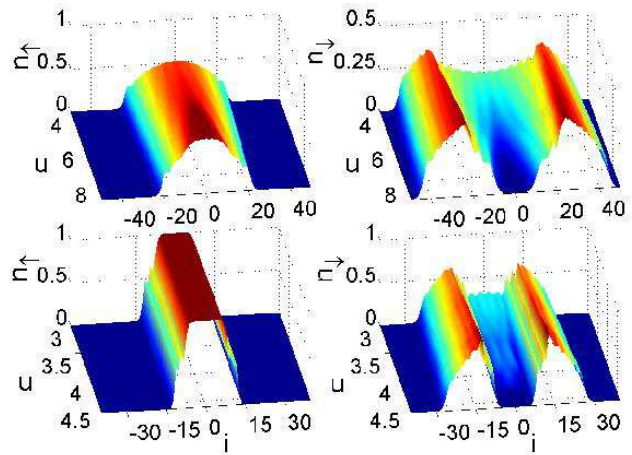}}
\caption{(Color online) The 3D plot for the spin-up and spin-down densities as functions of the site $i$ and the interaction strength $u$. The parameters chosen are, $N_f=40$, $P=0$, and $v_\downarrow=0.006$. The upper panels is for fixed $\gamma=3$, locating at the left of the minimum point in the phase boundary with $u\in[4,9]$. In this case, the spin-up densities are always smaller than one during the demixing process.
The Bottom panels is for fixed $\gamma=5$, locating at the right of the minimum point in the phase boundary with $u\in[3,4.6]$.
In this case, the spin-up densities form an insulating core during the demixing process, which leads to an extra energy needed to deplete completely
the spin-down atoms.}
\label{critical}       
\end{center}
\end{figure}

To have a overview on the transformation of the system from PM to PS region under the interplay of the ratio of the external potential and the repulsive interaction, we show the $\gamma-u$ state diagram in Fig. \ref{fig:phase} for different parameter systems. For the unpolarized systems, three different $v_\downarrow=0.001, 0.003$, and $0.006$ are shown. A polarized system of $v_\downarrow=0.001$ and $P=0.5$ is also included. As mentioned in the Ref.~\cite{pra-spin}, there exists a critical repulsive interaction strength ($u_c$) above which the demixing evolves. All the critical points form the phase boundary which distinguishes the PM phase from the PS one. For the unpolarized system, the different $v_\downarrow$ give very similar phase boundaries. For the systems of the same ratio $\gamma$, the demixing of the polarized system is easier due to more spin-up atoms and less spin-down atoms in the trap. Thus, smaller $u_c$ is observed for the onset of the phase separation.

By carefully investigating the different phase boundaries, we find that there is a minimum point in each of the phase boundaries, labeled by $(\gamma_0, u_0)$, which is closely related to whether the insulating core of spin-up atoms is formed in the trap center. To understand this minimum point, we show in Fig.~\ref{critical} the 3D plot of spin-up and spin-down densities for the system of $N_f=40$, $P=0$, and $v_\downarrow=0.006$. $\gamma$ is fixed to be 3 and 5 in the upper and bottom panels, corresponding to the left and right part of the minimum point [i.e., the third point: 3 $(\gamma_0\approx 4, u_0\approx 4.32)$ in Fig.~\ref {fig:phase}] in the phase boundary, respectively. We check the spin-up and spin-down densities by varying the interaction strength. We find that the local spin-up densities for $\gamma=3$ are always smaller than one ($n_{i\uparrow}<1$) during the demixing process, i.e., there is no insulating plateau. While the spin-up densities for $\gamma=5$ form an insulating core ($n_{i\uparrow}=1$) during the demixing process. From the above analysis we know that, the larger the $\gamma$, i.e., the tighter the confinement, the easier is to deplete the spin-down atoms. As a result, smaller $u_c$ is needed to achieve PS. This is the reason why the phase boundary firstly goes down when increasing $\gamma$. However, when the insulating plateau appears, the depletion of the spin-down atoms becomes more difficult and stronger on-site interaction is needed to achieve PS. That is, larger $u_c$ is needed, which makes the phase boundary go up slightly.

\section{Conclusions}
\label{conclusions}
In this paper, we discuss the demixing of fermion mixtures in optical lattices under the presence of a spin-dependent external potential by means of the DMRG and LSDA methods. The effects of the ratio of the spin-up and spin-down dependent external potential $\gamma$ and the repulsive interaction $u$ on the phases of the system are analyzed in details. The competition between the spin-dependent potential energy and the repulsive interaction energy is responsible for the phase separation. The external potential tends to push the atoms to the center of the system, while $u$ tends to delocalize atoms.
The $\gamma-u$ state diagram is obtained in which the phase boundary distinguishes the PM from the PS phases. A minimum point in each of the phase boundaries is observed. In the left part of this point, the local spin-up densities $n_{i\uparrow}$ are always smaller than one which makes the depletion of the spin-up comparatively easier. While in the right part of this point, the spin-up densities form an insulating core ($n_{i\uparrow}=1$) during the demixing process. The insulating plateau makes the depletion of the spin-down atoms more difficult and as a result, the phase boundary in this case goes up slightly.

The detailed state diagram found in this paper is helpful in choosing the optimal parameters to achieve demixing of the constituents and on the other hand useful in controlling the efficiency of a sympathetic cooling for the mixtures.

\section{Acknowledgements}
\label{Acknowledgements}
This work was supported by Zhejiang Provincial NSFC under Grant no. R6110175 and NSFC under Grants no. 10974181 and no. 11174253.



\begin{thebibliography}{}
\bibitem{footnote}
   When defining the PS region by requiring more extended local occupation in the trap center vanishing, the
   PS region is slightly diminished. We have thus checked the case
   by defining $n_{i\downarrow}\lesssim 10^{-3}$ with $i$ ranging over
   6 core sites. We found that the phase boundary does not change qualitatively in this case.
\end{thebibliography}
\end{document}